\documentclass[12pt]{article}
\usepackage[utf8]{inputenc}
\usepackage{authblk}
\usepackage{color}
\usepackage{cite}
\usepackage{xcolor}
\usepackage{graphicx}
\usepackage[colorlinks = true,
            linkcolor = blue,
            urlcolor  = blue,
            citecolor = blue,
            anchorcolor = blue]{hyperref}
\usepackage{ulem}
\usepackage[a4paper]{geometry}
\title{Will Gravitational Waves Discover the First Extra-Galactic Planetary System?}

\author[1]{Camilla Danielski}
\author[2,*]{Nicola Tamanini}

\affil[1]{UCL CSED, Atlas Building,  Office G23 - 25, Fermi Avenue, Harwell Campus, Didcot, OX11 0QR, United Kingdom. \textit{camilla.danielski@cea.fr}}
\affil[2]{Max-Planck-Institut für Gravitationsphysik, Albert-Einstein-Institut, Am Mühlenberg 1, 14476 Potsdam-Golm, Germany. \textit{nicola.tamanini@aei.mpg.de}}
\affil[*] {\small{Corresponding author}}

\date{\today}

\begin{document}

\maketitle

\begin{center}
    \textbf{This essay received an honorable mention in the Gravity Research Foundation 2020 Awards for Essays on Gravitation}
\end{center}

\begin{abstract}
    Gravitational waves have opened a new observational window through which some of the most exotic objects in the Universe, as well as some of the secrets of gravitation itself, can now be revealed.
    Among all these new discoveries, we recently demonstrated [N.~Tamanini \& C.~Danielski, \textit{Nat.~Astron.}, 3(9):858–866 (2019)] that space-based gravitational wave observations will have the potential to detect a new population of massive circumbinary exoplanets everywhere inside our Galaxy.
    In this essay we argue that these circumbinary planetary systems can also be detected outside the Milky Way, in particular within its satellite galaxies.
    Space-based gravitational wave observations might thus constitute the mean to detect the first extra-galactic planetary system, a target beyond the reach of standard electromagnetic searches.
\end{abstract}

\newpage 
In the history of mankind there have been scientific discoveries which have transformed our perception of the world and revolutionised our knowledge of the Universe.
Some of these achievements represent ``a giant leap for mankind", quoting Niel Armstrong when he first stepped on the Moon in 1969. Since then numerous keystone discoveries were made, in particular the detection of the first so-called \textit{exoplanet} \cite{51Peg}, i.e.~a planet outside the Solar System, and the first direct detection of \textit{gravitational waves} (GWs) \cite{Abbott:2016blz}.
The implications of these findings were so important in the context of contemporary science, that two out of the last three Nobel prizes for physics were awarded to commemorate these discoveries.
Nowadays exoplanets and GWs are experiencing an exceptionally fast development, and are possibly the two subjects in the field of astrophysics that most spark the imagination and expectations of people. 


On the one hand, the information provided by more than four thousands discovered exoplanets changed the place that our Solar System occupies in the galactic context. Before realising the huge diversity 
that the exoplanet population could offer, the Solar System used to be our reference case to develop planet formation theories.  Now, the picture has been revolutionized: our Solar System is only one example, out of thousands, of all possible outcomes of the planetary formation and migration processes. Last but not least we are closer than ever to detect the Earth's twin in the habitable zone \cite{Gilbert2020, Rodriguez2020}. 

On the other hand, since the first direct observations of GWs in 2015 \cite{Abbott:2016blz} ended a long experimental quest and opened a new gravitational window onto the Universe, GW observations began unlocking some of the outstanding secrets of general relativity.
They enable us to probe gravity at the highest energies and at its most dynamical environments, and provide us with insights on some of the most exotic objects populating the Universe, above all black holes and neutron stars \cite{Barack:2018yly,LIGOScientific:2018jsj,LIGOScientific:2018mvr,LIGOScientific:2019fpa,Abbott:2020uma}.
One day GWs might even let us peer into the very beginning of the Universe, just instants after the Big Bang \cite{Krauss:2014sua,Caprini:2018mtu}.

The future of GWs looks brighter than ever.
Space-based detectors, such as the \textit{Laser Interferometer Space Antenna} (LISA) \cite{Audley:2017drz},
will explore a new portion of the GW discovery landscape populated by new and different sources that emit GWs at  frequencies ($\sim$mHz) lower than those measured by Earth-based detectors ($\sim$10-1000 Hz).
Among this plethora of sources, LISA will be able to individually resolve tens of thousands of compact double white dwarf (DWD) binaries within and around the Milky Way \cite{2017MNRAS.470.1894K,Lamberts:2019nyk}.
Binary white dwarfs are the objects that will allow us to expand our understanding on planetary formation and planetary evolution, since they constitute a mean for detecting massive circumbinary exoplanets, as we recently demonstrated \cite{Tamanini:2019awb,LISAexoGW2}.

\begin{figure}
    \includegraphics[width=\textwidth]{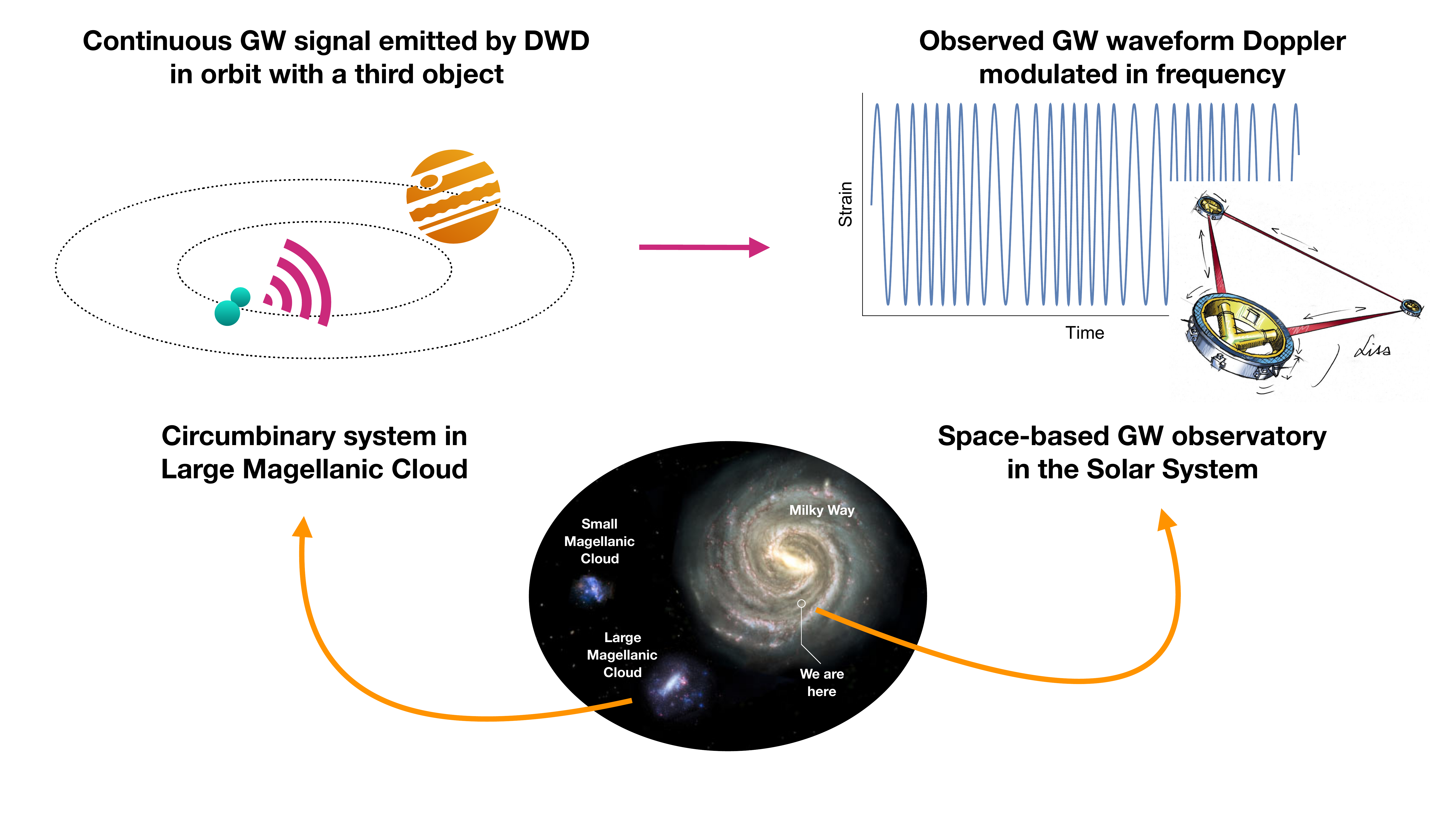}
    \caption{Simple schematic representation on how a space-based GW observatory such as LISA can observe circumbinary exoplanets outside the Milky Way. More details in the text.
    Credits: ESA for LISA cartoon; Nina McCurdy / Nick Risinger / NASA for image of galaxies.
    }
    \label{fig:cool}
\end{figure}
The idea of combining exoplanetary searches with GW observations might seem far-fetching, but the actual physical mechanism through which this can happen is surprisingly simple (Fig.~\ref{fig:cool}).
Galactic DWDs emitting in the LISA frequency band will merge no sooner than thousands of years and consequently their emitted GWs constitute a continuous, almost monochromatic (i.e.~with constant frequency) signal that LISA will be able to observe throughout its entire mission lifetime, which could be as long as 10 years.
Given this continued period of observation, LISA will be able to measure the GW frequency of these signals with extreme accuracy, implying that any small deviations from the almost monochromatic behaviour will be easy to detect.
Such a frequency shift can be induced by the gravitational attraction of a third object orbiting the DWDs \cite{Robson:2018svj}, for example a circumbinary planet.
This third body will force the DWD center of mass to move on a small Keplerian orbit and consequently to imprint a periodic Doppler modulation on the frequency of the produced GW signal \cite{Tamanini:2019awb} (cf.~Fig.~\ref{fig:cool}, top-right).
By detecting and characterizing this Doppler modulation, LISA will have the potential to estimate the period and the mass of the third object if its mass is comparable or heavier than Jupiter's and its separation from the DWD is less than 10 au \cite{Tamanini:2019awb}.
Our recent estimates predict that LISA could detect up to few hundreds exoplanets and a larger number of more massive circumbinary objects \cite{LISAexoGW2}. 

Such a population of exoplanets, which will be probed for the first time, can be detected all over the Milky Way, overcoming the horizon limitation of current and upcoming electromagnetic exoplanetary surveys, restricted to observe in the Solar neighbourhood. 
Space-based GW astronomy will thus be the only technique that will allow us to unveil the true Galactic planetary population in the next two decades.

Even more interesting is that DWD observations with space-based GW detectors are not limited to the Milky Way. LISA has the sensitivity to detect compact stellar binaries up to the limit of the local group, and in particular in nearby galaxies \cite{Korol2018}.
Recent estimates show that up to few hundreds DWDs might be detected in the Large Magellanic Cloud (LMC) \cite{Korol:2020lpq}, constituting the closest and more massive satellite galaxy of the Milky Way (cf.~Fig.~\ref{fig:cool}).
Besides yielding information on the number and features of Milky Way satellite galaxies \cite{Korol:2020lpq,2020arXiv200210465R}, these observations could also deliver the detection of the first extra-galactic planetary systems, as we argue in what follows.

So far no gravitationally bound exoplanets have ever been detected outside our Galaxy.
As mentioned above current electromagnetic techniques are limited to observations in the Solar neighbourhood, with the exception of microlensing measurements that can detect as far away as the Galactic Buldge.
Microlensing is more sensitive towards stellar crowded regions, and has already been used to detect free-floating, i.e.~unbound, planets in other galaxies \cite{DaiGuerras2018}.
Needless to say that planets bound to a star, or to a binary star, are more interesting from a scientific point of view, not only because they provide information on the formation and evolution of whole planetary systems, but also because stellar irradiation is an important aspect in the evolution of a planet, and a basic criteria for supporting life as we know it.

By following the same analysis and detectability conditions of \cite{Tamanini:2019awb}, and taking into account the results recently obtained in \cite{Korol:2020lpq}, here we explore the opportunities for the detection with GWs of circumbinary exoplanets in the LMC.
We simulate the response of LISA to the GWs signal emitted by an equal mass DWD placed in the LMC and orbited by a circumbinary planet at a separation of 1 au with edge-on orbital inclination.
By setting the total DWD mass to 0.5 $M_\odot$ (Solar mass) and its period to 5 min, in agreement with the fiducial extragalactic system considered in \cite{Korol:2020lpq}, a 13 M$_{\rm J}$ (Jupiter's mass) circumbinary planet will appear at the limit of detectability after 4 years of LISA mission operation, and it will be certainly detected with an extended 10 year mission. 
The same planet will be detected within 4 years for a higher frequency DWD with 3 min period, for which 10 years of LISA observations will even have the potential to detect a similar 5 M$_{\rm J}$ planet.
Given the fact that both circumbinary detections and extragalactic DWDs observed by LISA will be biased towards the higher possible DWD periods \cite{Tamanini:2019awb,Korol:2020lpq}, we might expect reasonable chances for the detection of an extragalactic circumbinary planetary system already with LISA, if their rates are similar to the more optimistic ones estimated for the Milky Way \cite{LISAexoGW2}.
In any case, even if LISA will not have the sensitivity required to detect such extra-galactic planets, a future LISA-like mission, with an improved sensitivity of at least one order of magnitude around the mHz frequency region, similarly to the envisaged $\mu$Ares \cite{Sesana:2019vho} or AMIGO \cite{Baibhav:2019rsa} space missions, should be able to spot a 4 M$_{\rm J}$ planet orbiting a 5 min period DWD with 10 years of observations.
A two order of magnitude improvements would allow for the detection of extra-galactic planets in the LMC with Jupiter-like mass.
Note that, according to our estimates, circumbinary objects with masses heavier than 13 M$_{\rm J}$, alias the deuterium burning limit which distinguishes planets from more massive bodies, will be even easier to detect. This will possibly lead to the extra-galactic observation of triple stellar systems or circumbinary brown dwarfs \cite{Robson:2018svj,LISAexoGW2}.
Our estimates above will considerably worsen if the third object appears with a much larger separation than 1 au, or with an orbital inclination far from edge-on, in which case only heavier circumbinary masses will be detectable.

The simple calculations reported above show that future GW space missions could deliver the first detection of an extra-galactic planetary system.
Besides confirming the existence of bound planets outside our own Galaxy, such a milestone observation would provide useful scientific insights on the nature of planetary populations in other galaxies.
The fact that space-based GW detectors will one day be able to detect circumbinary exoplanets in nearby galaxies, can thus be well considered an outstanding achievement, highlighting the ubiquitous multidisciplinary opportunities of GW astronomy.

\bibliographystyle{unsrt}
\bibliography{mybib.bib}

\end{document}